\def \SAIT #1 #2 {{\em Mem.\ Soc.\ Astron.\ It.\/} {\bf #1}, #2}
\def \MESS #1 #2 {{\em The Messenger\/} {\bf #1}, #2}
\def \ASTRNACH #1 #2 {{\em Astron. Nach.\/} {\bf #1}, #2}
\def \AAP #1 #2 {{\em Astron. Astrophys.\/} {\bf #1}, #2}
\def \AAL #1 #2 {{\em Astron. Astrophys. Lett.\/} {\bf #1}, L#2}
\def \AAR #1 #2 {{\em Astron. Astrophys. Rev.\/} {\bf #1}, #2}
\def \AAS #1 #2 {{\em Astron. Astrophys. Suppl. Ser.\/} {\bf #1}, #2}
\def \AJ #1 #2 {{\em Astron. J.\/} {\bf #1}, #2}
\def \ANNREV #1 #2 {{\em Ann. Rev. Astron. Astrophys.\/} {\bf #1}, #2}
\def \APJ #1 #2 {{\em Astrophys. J.\/} {\bf #1}, #2}
\def \APJL #1 #2 {{\em Astrophys. J. Lett.\/} {\bf #1}, L#2}
\def \APJS #1 #2 {{\em Astrophys. J. Suppl.\/} {\bf #1}, #2}
\def \APSS #1 #2 {{\em Astrophys. Space Sci.\/} {\bf #1}, #2}
\def \ASR #1 #2 {{\em Adv. Space Res.\/} {\bf #1}, #2}
\def \BAIC #1 #2 {{\em Bull. Astron. Inst. Czechosl.\/} {\bf #1}, #2}
\def \JSQRT #1 #2 {{\em J. Quant. Spectrosc. Radiat. Transfer\/} {\bf #1}, #2}
\def \MN #1 #2 {{\em Mon. Not. R. Astr. Soc.\/} {\bf #1}, #2}
\def \MEM #1 #2 {{\em Mem. R. Astr. Soc.\/} {\bf #1}, #2}
\def \PLR #1 #2 {{\em Phys. Lett. Rev.\/} {\bf #1}, #2}
\def \PASJ #1 #2 {{\em Publ. Astron. Soc. Japan\/} {\bf #1}, #2}
\def \PASP #1 #2 {{\em Publ. Astr. Soc. Pacific\/} {\bf #1}, #2}
\def \NAT #1 #2 {{\em Nature\/} {\bf #1}, #2}

\documentstyle{memsait}
\input epsf.sty
\begin{opening}
\title{NATURE OF THE SOFT GAMMA REPEATERS AND ANOMALOUS X-RAY PULSARS} 
\author{CHRISTOPHER THOMPSON$^1$}
\institute{$^1$
Canadian Institute for Theoretical Astrophysics,
60 St. George St., Toronto ON, M5S 3H8
}
\date{} 
\end{opening}

\begin{document}

\oddpagefooter{}{}{} 
\evenpagefooter{}{}{} 
\
\bigskip

\begin{abstract}
I summarize recent developments in the magnetar model of the
Soft Gamma Repeaters and Anomalous X-ray Pulsars, give a critical
inventory of alternative models for the AXPs, and outline the
improved diagnostics expected from present observational efforts.
\end{abstract}

\section{Introduction}
The Soft Gamma Repeaters are young neutron stars detected through their
brilliant flashes of X-rays and gamma-rays.  Strong physical arguments
point to the presence of $\sim 10^{14}-10^{15}$ G magnetic fields in these
sources.  (A detailed analysis of 27 August 1998 flare data is given
in Feroci et al. 2001, and of the flare physics in Thompson \& Duncan 2001.)
The Anomalous X-ray Pulsars have gradually emerged as a class of neutron
stars with moderate luminosities
($L_X \sim 3\times 10^{34}-10^{36}$ erg s$^{-1}$) and a narrow range of spin
periods ($P \sim 6-12$ s).  The AXPs are persistently spinning down,
some are observed to vary in brightness, and about half are convincingly
associated with young supernova remants.  Remarkably, the SGRs in their
quiescent states have similar luminosities and spins to the AXPs, and
are spinning down even more rapidly.  This overlap between the
two classes of sources within a three-dimensional parameter space
(in spite of the very different detection methods)
motivated the suggestion that both are isolated neutron stars powered
by the same energy source:  the decay of an ultrastrong magnetic field
(Thompson \& Duncan 1993, 1996).  Recent data suggest two
objects as possible `missing links' between the two classes:
SGR 0525-66, which has been dormant as a burst source since 1983
and has a soft AXP-like power-law spectrum (Kulkarni et al. 2000);
and the anomalous pulsar 1E 1048.1-5937, which has a harder persistent
X-ray spectrum than SGR 0525-66, and a noisier spindown torque
than the other AXPs (Kaspi et al. 2000).
Nonetheless, the observed properties of the AXPs are less constraining
than those of the SGRs, and have inspired a number of models over the
last decade.

\section{Models of the Anomalous X-ray Pulsars}
The AXPs were first defined as a group\footnote{The name `Anomalous X-ray
Pulsar' was coined by Duncan \& Thompson (1995) in a
conference presentation.} by Mereghetti \& Stella (1995).
Theoretical models for the AXPs must explain several
peculiar properties of these sources:  i) the absence of a modulation of
the X-ray pulse frequency by binary motion, which is consistent with
a low mass helium star, or a very low mass degenerate dwarf companion
(Mereghetti, Israel, \& Stella 1998);  ii) low optical fluxes,
$L_{\rm opt}/L_X \leq 10^{-3}$ (Hulleman et al. 2000a,b);
iii) persistent spindown, $\dot P > 0$, much larger than the $\dot P$
of radio pulsars, and young characteristic ages
$P/\dot P = 4\times 10^3-4\times 10^5$ yrs; iv) a wide range of torque noise,
with some AXPs spinning down almost as quietly as radio pulsars
(Kaspi et al. 1999, 2001);  iv) spindown luminosities
$I\Omega\dot\Omega \sim 10^{32}-10^{33}$ erg s$^{-1}$, some 2-3 orders
of magnitude smaller than the X-ray luminosities
$L_X \sim 3\times 10^{34}-10^{36}$ erg s$^{-1}$; and v) the localisation
of at least 3 AXPs close to the geometrical centers of young
supernova remnants (e.g. Gaensler et al. 2001).  See the paper by
Israel, Mereghetti, \& Stella (these proceedings) for a more complete review.

Here is an inventory of the models for the AXPs,
listed in the chronological order in which they were invented.
I will indulge in one editorial comment at the beginning:  In constructing
models, there has sometimes been a tendency to equate the `symptoms' with the
underlying `disease'.  Before the splendid phase-connected
spin measurements by Kaspi et al. (of the AXPs) and Woods et al. (of the
the SGRs) the only neutron stars showing large torque noise were accreting
X-ray binaries and a few young glitching pulsars (such as Vela).  The most
conservative approach would be to fit the AXP/SGR measurements into
one of these classes, and this indeed has been the strongest motivation for
the accretion-based torque models.  There is, however, a fairly direct
argument that we are observing a {\it new} mechanism for
torque noise in neutron stars.  The torque acting on an isolated neutron
star is a direct measure of the current flowing through its outer
magnetosphere; but the magnetic fields of the SGR sources are time variable
(giving rise to the bright X-ray bursts), and so one can also expect
the current to be time-variable.  The only leap one must make here, with
regard to the AXP sources, is to postulate that the magnetic field (and
the crust which anchors it) are capable of more gradual deformations,
in addition to the sudden yields and fractures which appear to
trigger SGR flares.  Nonetheless, the patterns of torque noise in
the SGRs and AXPs are turning out to be sufficiently similar,
that any torque model should be able to accomodate both classes of sources.

i) {\it Massive white dwarfs, formed in white dwarf mergers, now spinning
down by magnetic dipole torques.}  Paczy\'nski (1990) suggested this
model for 1E 2259+586, which was recognized early on as a peculiar object
(Fahlman \& Gregory 1981).  The spindown luminosity
of a white dwarf is much larger, at fixed $P$ and $\dot P$, than that
of a neutron star, by the ratio of moments of inertia
$I_{\rm wd}/I_{\rm ns} \sim
0.3\,(M_{\rm wd}/M_{\rm ns})\,(R_{\rm wd}/R_{\rm ns})^2 > 3\times 10^4$.
The required polar magnetic field is
\begin{equation}
B_{\rm pole} R_\star^2 = {9\times 10^{26}\over \sin\alpha}\,
\left({P\dot P\over 10^{-10}~{\rm s}}\right)^{1/2}\,
\left({M_\star\over M_\odot}\right)^{1/2}\,
\left({I_\star\over M_\star R_\star^2}\right)^{1/2}\;\;\;{\rm G-cm^2},
\end{equation}
from the standard dipole formula with inclination angle $\alpha$.
King et al. (2001) have noted that a binary merger is a promising
formation scenario for the white dwarf RE J0317-853
($B \sim 4-8\times 10^8$ G), but its $725$-s spin is very slow
for the AXPs (albiet fast for a white dwarf).  The questions which
hang over the white dwarf interpretation of the AXPs are fairly obvious:
why is the radiative area deduced from the thermal component of AXP
emission comparable to that of a neutron star (Perna et al. 2001)?
Why are half the AXPs situated close to the centers of supernova
remnants, whose kinetic energy greatly exceeds the rotational energy
of the white dwarf?  Why are the white dwarfs visible only while young,
and why the narrow range of spin periods?

ii) {\it Magnetars:  neutron stars with decaying $\sim 10^{14}-10^{15}$
G magnetic fields.}  Thompson \& Duncan (1993, 1996) drew a connection
between 1E 2259+586 and the SGR 0525-66, based on their similar spin periods,
persistent $L_X$, and residence in $\sim 10^4$-yr old SNR.
The rapid spindown of the AXPs has a simple interpretation as magnetic dipole
breaking, modified by variable magnetospheric currents flowing along
the decaying magnetic field (Thompson, Lyutikov, \& Kulkarni 2001).
A variable X-ray flux also points to active field decay in some sources.
The most interesting argument against accretion as the energy source
of AXP emission is indirect and comes from the SGR sources:
in particular, SGR 1900+14 was detected as a persistent
source within $\sim 10^3$ s of the August 27 giant flare, and yet the
radiative momentum of the flare would excavate any accretion disk and
suppress accretion for a much longer period (Thompson et al. 2000).

The limited duration of AXP/SGR activity is consistent with the
freeze-out of ambipolar diffusion of the magnetic field through the
core of a neutron star after surface photon cooling dominates (Thompson
\& Duncan 1996; Heyl \& Kulkarni 1998).  The characteristic timescale
is $\sim 10^5$ yrs for an iron surface composition, and $\sim 10^4$ yrs
for H/He.  However,
the observed range of spin periods is surprisingly narrow, given the
wide range of characteristic ages exhibited by magnetar candidates:
from $P/\dot P = 1-3,000$ yrs for the spinning down SGRs 1806-20
and 1900+14 (Kouveliotou et al. 1998; 1999) to $P/\dot P = 4\times 10^5$ yrs
for 1E 2259+586.  This points, in a model-independent manner, to
a {\it decay} in the torque acting on 1E 2259+586, from its historic average
(Thompson et al. 2000).  The measured spins are remarkably close
to the upward extrapolation of the radio pulsar death line in the
$\dot P-P$ plane.  If the magnetar model were to explain them,
it would probably be on the basis of a connection between a pair-loaded
current, and the stability of a non-potential magnetic field outside the star.

iii) {\it Low Mass X-ray Binaries.}  In defining the AXPs,
Mereghetti \& Stella (1985) suggested that they are accreting neutron
stars with very low mass companions.  They were motivated by
the similarity between the spin periods of the AXPs and the equilibrium
spin period $10\,(B/3\times 10^{11}~{\rm G})^{6/7}\,
(L_X/10^{35}~{\rm erg~s^{-1}})^{-3/7}$ s of an accreting neutron star
whose magnetic field is typical of (in fact, slightly weaker than) a young
radio pulsar.  In addition, the known LMXB 4U 1626-67 has a spin and
X-ray luminosity ($P = 7.7$ s, $L_X \sim 10^{36}$ erg s$^{-1}$)
in the AXP range.  The X-ray spectrum of 4U 1626-67 is, however, much
harder than that of any AXP, and there is an optical counterpart
(Chakrabarty 1998).  This
system also has a feature which is natural in any LMXB model for the AXPs,
but is not observed in those sources as presently defined:  4U 1626-67
has exhibited intervals of rapid spin-up as well as spin-down,
which is expected as the source is old.  The long evolutionary timescale
leads more generally to a prediction of too many AXP sources, given
the probable association of 3 AXPs with $\sim 10^4$-yr old supernova
remnants:  a binary with a companion of mass
$M_2 \sim 0.1\,M_\odot$, undergoing Roche-lobe overflow with a luminosity
$L_X = GM_{\rm NS}\dot M/R_{\rm NS} = 10^{35}$ erg s$^{-1}$, would
evolve on the much longer timescale $M_2/\dot M \sim 1\times 10^{10}
(M_2/0.1\,M_\odot)\,(L_X/10^{35}~{\rm erg~s^{-1}})^{-1}$ yrs.
Overall, it is not surprising that at least one of the large number
of accreting X-ray binaries should happen to overlap the AXP parameter space.

iv) {\it Neutron Stars with Fossil Disks.}  Some neutron stars may
capture a quasi-Keplerian disk after their formation, a scenario which
has received renewed interested as a model for the AXPs (Van Paradijs,
Taam, \& Van den Heuvel 1995; Chatterjee, Hernquist, \& Narayan 2000).
As was realized after the discovery of the first pulsar planetary system,
several evolutionary pathways could result in the formation of such a disk.
The accretion rate will generically drop off with time, as the orbiting
material spreads outward.  If the spin of the neutron star is able to
track the equilibrium spin period $P_{\rm eq} \propto \dot M^{-3/7}$
(where the corotation radius is close to the Alfv\'en radius) then its
spin will decrease with time.  Li (1999) has noted that in
some sources (e.g. 1E 1048.1-5937) the time to return to the equilibrium
spin is much longer than the measured characteristic age $\sim P/2\dot P$,
unless the X-ray emission is significantly beamed and $\dot M$ is
underestimated.   This is not a concern in the case of an old accreting
binary with fluctuating $\dot P$, but it is if the source is only
$\sim 10^4$ years old.

Several questions hang over this model:  i) Why are the AXPs so similar
to the SGRs in their quiesecent states?  It has been suggested that
the torque is modified by accretion effects in the SGR sources
(Marsden et al. 2001), but no cogent connection between accretion and
bursting activity has been given.  ii) An upper age limit of
$\sim 10^4-10^5$ yrs could result from the dissipation of the disk
(e.g. Chatterjee \& Hernquist 2000), but a narrow range of $P$
would seem to require a narrow range of initial disk masses.
iii) Optical/IR emission from the portion of the disk exposed to
the central X-ray source.  The possible detection of an optical counterpart
to the AXP 4U 0142+61 with (de-reddened) flux $\sim 10^{-3}$
of the X-ray flux (Hulleman et al. 2000b) conflicts with the expectation
of $L_{\rm opt}/L_X\sim 10^{-2}$ (Perna, Hernquist, \& Narayan 2000),
and the spectrum is much redder than is typical of a LMXB.
iv) A basic theoretical question is whether an isolated neutron star,
surrounded by a centrifugally supported envelope, is able to lose angular
momentum through a propeller and approach corotation with a weak
accretion flow.   Whether this
is possible depends on the angular momentum carried away by each gram of
ejected material, the initial spin, and the disposition of the
surrounding material (in a disk vs. an extended envelope).  Calculations
which answer in the affirmative have assumed a disk geometry and a
fast initial spin, but have allowed the ejected material to be spun
up to corotation with the star -- the most optimistic prescription for
the propeller (Chatterjee et al. 2000).  The critical dipole field,
above which the rotation is slowed down to allow accretion, rises
from $\sim 10^{13}$ G to more than $10^{15}$ G if one makes a different
(weaker) prescription for the propeller:  that the angular momentum
per gram is that of a Keplerian orbit at the magnetospheric boundary.

%
%

\section{Magnetars:  Recent Developments}
The magnetar model for the Soft Gamma Repeater sources and
the Anomalous X-ray Pulsars has been reviewed recently (Thompson 2000),
and so I will list a few recent theoretical highlights here.  A case
can be made that we now have a better understanding of the radiative
mechanism and the nature of the `engine' in SGR bursts, than has yet
been obtained for the more distant cosmological GRBs.

i) {\it Direct Evidence for Trapped Fireballs in SGR Giant Flares.}
After the initial $\sim 0.5$-s fireball, followed by a
$\sim 40$-s phase of irregular pulsations, the X-ray flux in the
27 August giant flare maintained a smooth decline until an abrupt
termination at $\simeq 380$ s.  After averaging over the large-amplitude
$5.16$-s pulsations, the light curve is well fit by the contracting
surface of a trapped fireball, $L_X(t) \simeq L_X(0)(1-t/t_{\rm evap})^3$
(Feroci et al. 2001).  Because the shape of the lightcurve is changed by
neutrino cooling (which is strongly temperature-dependent), the diameter
of the confined $e^\pm$ plasma cannot be much less than 10 km.  This
leads to a strict lower bound $m/R_{\rm NS}^3  \sim  3\times 10^{13}$ G
on the dipole magnetic moment $m$ of the confining field (Thompson \&
Duncan 2001).  As the X-ray flux declines, the thermal component of
the X-ray spectrum maintains a constant temperature $\simeq 11$ keV
(black body value: Mazets et al. 1999; Feroci et al. 2000), identical
to the temperature at which photon splitting freezes out
(Thompson \& Duncan 1995).

Independent evidence for a trapped fireball comes from a less energetic
burst emitted by SGR 1900+14 two days after the 27 August giant flare.
The main burst was followed by a faint
pulsating tail lasting at least $\sim 10^3$ s, during which the
excess flux decreases as $\sim t^{-0.6}$ (Ibrahim et al. 2001).
The high temperature of this tail implies a radiative area only
$\sim 1$ percent that of a neutron star.  This area is consistent
with a trapped fireball of a small volume, but an energy density
similar to that attained in the much longer pulsating tail of the giant flare.
The energy released in this faint tail can be accounted for by
heating of the outer crust through $e^-$-captures, while it was compressed
under the fireball pressure (Ibrahim et al. 2001).

ii) {\it Thermal Stability of Magnetized Pair Plasmas, and Short SGR Bursts.}
The trapped fireball model requires that energy is released fast
enough into the magnetosphere to create a plasma in local thermodynamic
equilibrium.  A steady balance between heating and radiative cooling from
the (optically thick) plasma could be maintained if the rate of
injection of energy were less than $\sim 10^{42}\,
(R/10~{\rm km})$ erg s$^{-1}$ within a volume $R^3$ (Thompson \& Duncan 2001).
We know that LTE can be achieved in the giant flares, because their
peak luminosities exceed $\sim 10^{44}$ erg s$^{-1}$ during the initial
$\sim 0.5$-s transient.  Similarly, one infers that the injection was
fast enough to reach LTE in the main, bright
component of the 29 August burst.  Nonetheless, the question remains
open as to whether an LTE fireball actually forms in the much more common
short SGR bursts, which have a characteristic $\sim 0.1$-s duration
much shorter than the $3.5$-s duration of the 29 August burst
({G\"o\u{g}\"u\c{s}} et al. 2001).
In a steady state, the heated electrons cool primarily by Compton
upscattering O-mode photons (which maintain a Wien distribution and
have a large scattering cross-section $\sim \sigma_T$ in an ultra-strong
magnetic field).  The net result is an anti-correlation between hardness
and luminosity (Thompson \& Duncan 2001) which is qualitatively similar
to that observed in the short bursts ({G\"o\u{g}\"u\c{s}} et al. 2001).

iii) {\it Passively Cooling, Ultramagnetized Neutron Stars.}  The
transparency of the outer envelope of a neutron star is increased slightly
in the presence of a strong magnetic field, which led to the suggestion
that some AXPs may be passively cooling neutron stars without
active magnetic field transport (Heyl \& Hernquist 1997).  This model
is simple enough for detailed calculations of the angular and spectral
distributions of the cooling X-ray flux to be compared with AXP
data.  Gravitational bending of the photon trajectories tends to reduce
pulsed fractions (Psaltis, \"Ozel, \& DeDeo 2000).  In addition,
the spectrum will differ from a pure blackbody, as the result of the
strong asymmetry between the opacities of the two X-ray polarization modes,
which might explain some of the soft, high energy power-law tails
obtained in AXP spectral fits (\"Ozel 2001).  It is worth keeping in
mind that two effects will bias the number of observed sources toward
those with active magnetic field decay:  external currents are more
efficient at generating X-rays than is deep heating; and the lifetime of a
magnetar as a thermal X-ray source is lengthened, by an order of magnitude,
by the decaying field (Thompson \& Duncan 1995; Heyl \& Kulkarni 1998).

iv) {\it Magnetar Electrodynamics.}  The persistent emission of the
active SGRs has a hard, power-law spectrum and correlates directly with
X-ray burst activity (Woods et al. 2001).  A strong magnetic field will
heat the interior of a neutron star, and also drive electrical currents
through its exterior (Thompson \& Duncan 1996).
The lowest energy deformation of the crust, extending over a scale
of kilometers, involves a predominantly rotational motion, which will
twist up the external magnetic field.
Detailed force-free models of twisted, current-carrying magnetospheres
have been constructed by Thompson, Lyutikov \& Kulkarni (2001).  They
have the remarkable property that the optical depth to resonant cyclotron
scattering depends directly on the twist angle, but not on the mass,
charge, or radius of the scattering particle.  As a result, the smooth
pulse profile emerging in SGR 1900+14 after the 27 August giant
flare (Woods et al. 2001) has a simple explanation through the formation of
an extended scattering screen at the electron cyclotron resonance, which
for electrons sits at $R/R_{\rm NS} \simeq
10 (B_{\rm pole}/10^{14}~{\rm G})^{1/3}\,(\hbar\omega/{\rm keV})^{-1/3}$
in magnetar-strength fields.
In addition, multiple resonant scattering will generate an extended
high-energy tail in the persistent X-ray spectrum, and allow large pulse
fractions in AXP sources.  The field decreases more slowly
with radius than a pure dipole, but only slightly, so that the polar
surface fields inferred from the measured $\dot P$ and $P$ of the SGR
sources are reduced by a factor $3-10$ compared with respect to
the classical magnetic dipole model.

\section{Observational Tests and their Theoretical Implications}

In the next few years, the following observational efforts can be
expected to throw further light on the true nature of the SGR/AXP sources.

i) {\it Optical counterparts to the SGRs and AXPs.}  Besides the possible
detection $L_{\rm opt}/L_X \sim 10^{-3}$ for 4U 0142+61 (with a
peculiar spectrum: Hulleman et al. 2000), there is an interesting upper
limit for 1E 2259+586 (Hulleman et al. 2000a) and SGR 0525-66
(Kaplan et al. 2001).   High extinctions may limit further progress on
the SGR sources, but the prospects are good for improving optical
identifications of the AXPs by making use of improved X-ray
localizations.  Strong constraints on accretion will be obtained --
either through flux limits; or from the spectral distribution of the measured
optical/IR emission, combined with the cross-correlation between the X-ray
and optical fluxes.

ii) {\it Connection between the SGR/AXPs and nearby cooling neutron stars.}
The SGRs and AXPs fade as X-ray sources after $\sim 10^4-10^5$ yrs, but
little is known about the further evolution of their X-ray flux and
spectrum.  The nearby soft X-ray pulsars RX J0420.0-5022, RX J0720.4-3125,
and RBS1223 have similar spin periods (22.7, 8.37, and 5.2 s) to the
SGRs/AXPs, but much lower X-ray luminosities.  It is difficult to make
unambiguous predictions about the field decay occuring in a magnetar
after it begins photon cooling, because the heating quickly becomes
dominated by the crustal magnetic field.  Measurements of spindown
in these sources will provide crucial input to the physical modelling.

iii) {\it Ion cyclotron features:  emission vs. absorption?}  Detailed
calculations of cooling neutron star atmospheres in very strong
magnetic fields ($B = 10^{14}-10^{15}$ G) show that broad absorption
features are possible at the fundamental ion cyclotron resonance
(Zane et al. 2001; Ho \& Lai 2001).  Nonetheless, the persistent
emission of the SGRs and some AXPs has a strong non-thermal component
which, if interpreted in terms of stationary magnetospheric currents,
implies a significant amount of surface heating -- much of which could be
re-radiated through an emission line (Thompson, Lyutikov, \& Kulkarni 2001).

iv) {\it Torque variations in the SGR/AXP sources.}  Our picture of
torque variations in the SGR and AXP sources is still murky:  the
level of torque noise is sometimes not much higher than radio
pulsars (Kaspi et al. 1999), but at other times the torque can
vary by at least a factor $\sim 2$ over periods of months to years
(Woods et al. 2000; Paul et al. 2000; Kaspi et al. 2000).  It appears
so far that variations in the persistent output (and bursting emission)
do not correlate directly with torque variations.  Whether there
is some hysterisis between the two, or whether they are completely
uncorrelated, has important implications for the physical mechanism
and can only be determined by redoubling the efforts at phase-connected
timing (Kaspi et al. 1999, 2000; Woods et al. 2000).  These monitoring
campaigns are especially important if, as argued above, the variable
torque of the SGRs and AXPs is a distinct physical phenomenon from
that encountered in accreting binaries or glitching radio pulsars.

\acknowledgements
I thank Vicky Kaspi, Shri Kulkarni, and Pete Woods for conversations.
Financial support was provided by NASA (NAG 5-3100), the NSERC of Canada,
and the Alfred P. Sloan foundation.


\end{document}